\documentclass[aps,prb,twocolumn,groupedaddress,showpacs]{revtex4}
\usepackage{graphicx}
\usepackage{isolatin1}
\newcommand{\myscalebox}[1]{\scalebox{0.42}[0.42]{#1}}
\newcommand{\myscaleboxB}[1]{\scalebox{1.00}[1.00]{#1}}
\begin{document}
\title{Generating droplets in
  two-dimensional Ising spin glasses by using matching algorithms}
\author{A. K. Hartmann}
\affiliation{Institut f\"ur Theoretische Physik, Universit\"at G\"ottingen,
Tammannstrasse 1, 37077 G\"{o}ttingen, Germany}
\author{M. A. Moore}
\affiliation{Department of Physics and Astronomy, University of 
Manchester,
Manchester, M13 9PL, United Kingdom} 
\date{\today}
\begin{abstract}
We study the behavior of droplets
 for two dimensional Ising spin glasses with Gaussian
interactions. We use an exact matching algorithm which enables
study of systems with linear dimension $L$ up to 240, which is larger
 than is possible with
other approaches. But the method only allows certain classes of droplets
 to be generated. We study {\em single-bond}, {\em cross} and a category of 
{\em fixed  volume} droplets as well as first excitations. By comparison with
similar or equivalent droplets generated in previous works, the
advantages but also the limitations of this approach are revealed.
In particular we have studied the scaling behavior of the
 droplet energies and
droplet sizes. In most cases, a crossover of the data can be observed
such that for large sizes the behavior is compatible with the
one-exponent scenario of the droplet theory. Only for the case of
first excitations, no clear conclusion can be reached, probably because even
with the matching approach the accessible system sizes are still too small.
\end{abstract}
\maketitle

\section{Introduction}

Ising spin glasses are amongst the most-frequently studied systems in
statistical physics \cite{reviewSG}. However, despite more than two
decades of intensive research, many properties of spin glasses,
especially in finite dimensions, are still not well understood.
For two-dimensional Ising spin glasses it is now widely accepted that no
ordered phase for finite temperatures exists 
\cite{rieger1996,kawashima1997,stiff2d,houdayer2001,carter2002}.
For $d=2$ the behavior is usually described by a  zero-temperature  
droplet scaling (DS) approach \cite{McM,BM,FH}.

 The droplet picture assumes that the
low-temperature behavior is  
governed by droplet-like excitations, where
 excitations of linear spatial extent $l$ typically cost an energy
of order $l^{\theta}$. These excitations are
expected to be compact and their surface 
has a fractal dimension $d_s<d$, where $d$ is the space dimension.
  Furthermore it is usually
assumed that the scaling behavior of the 
energy $\delta E$ of different types of excitations,
e.g. droplets and domain walls, induced by changing  boundary
conditions, are described by the same exponent $\theta$. 

The value of
$\theta$ for domain walls in systems with 
Gaussian disorder has been determined by several studies 
\cite{bray1984,mcmillan1984b,rieger1996,palassini1999a,stiff2d,carter2002}
with results close to the most recent result \cite{aspect-ratio}
$\theta=-0.287(2)$. For simplicity, we will assume
$\theta=-0.29$ from now on.
On the other hand, in a previous study of droplets 
by Kawashima and Aoki \cite{KA}
using Monte Carlo
simulations of moderate sized systems, a different exponent
$\theta=-0.45(5)$ was found. 
Note that droplets can be generated
by using several methods, and the resulting scaling behavior may in principle 
depend on the
droplet type considered.
By using  exact ground-state (GS) algorithms
\cite{opt-phys2001}, so called {\em matching algorithms} (see
Sec. \ref{sec:algorithms}), one can study much larger system sizes than
 is possible with other approaches. On the
other hand matching algorithms are less flexible than Monte Carlo
methods, and not all types of droplets can be generated using them.
But recently, through the application of an extended matching algorithm
\cite{droplets},
we were able to generate the same type of droplets as in Ref. \onlinecite{KA}
but for much larger systems. For large systems, the
value $\theta=-0.29$ of the  exponent was again
found. To be precise, the scaling of the droplet energies
exhibited a crossover of the form 
\begin{equation}
\Delta E=Al^{\theta}+Bl^{-\omega}.
\label{basic}
\end{equation} with $\theta=-0.29$, $\omega=1$, while
for small sizes, the behavior was compatible with an apparent exponent
of $\theta^\prime=-0.47(1)$. 
Hence the predictions of
DS have been verified for this type of droplet, 
simply by studying larger system
sizes.

Recently, the same crossover phenomenon was sought by Berthier
and Young \cite{berthier2003} when studying droplets, obtained by a
Monte Carlo minimization in a system with size $L=64$
while fixing the size of the droplets to either one value
or to a range of values. In both cases, for the largest droplets, the
scaling was determined by an exponent resembling the domain-wall
exponent, i.e. close to $\theta=-0.23$. But only for the second type
of droplets with a range of sizes
could a crossover phenomenon be observed, and then it only
occurred for very small droplets. For a third type of
droplet, no crossover and a scaling exponent $\theta=-0.32$
close to the domain-wall value was found \cite{excited2d}.

On the other hand,  Picco, Ritort and Sales \cite{picco2001,picco2003}
have studied the lowest excitation in systems with 
sizes up to $L=16$ using a transfer matrix
approach. Through a scaling argument connecting 
 lowest excitations to droplets
(see Sec. \ref{sec:lowest}), 
their result implied $\theta=-0.47$.

In order try to clarify the above mentioned results and 
relate them to our previous
results \cite{droplets} 
 we study in this paper different 
types of droplets by applying again the matching
algorithm. This allows us to go to larger sizes than in the
studies where no matching algorithm was applied. On the other hand,
 the matching algorithm is less flexible and enables us to generate
 only certain types of droplets. Where differences in the scaling behavior
 arise amongst the various droplet types 
we shall attempt to understand their origin.

The Hamiltonian which is studied here is the usual  Ising
spin glass model:
\begin{equation}
{\mathcal H}=-\sum_{\langle i,j\rangle }J_{ij}S_iS_j,
\end{equation}
where the sites $i$ lie on the sites of a square lattice with $N=L^2$ sites,
$S_i=\pm1$, the $J_{ij}$ have a Gaussian distribution of zero mean and
unit variance and couple nearest-neighbor sites on the lattice.

In the next section, we will outline how the matching algorithm works
and how the droplets are generated. In the main four sections of this
paper, four different types of droplets are studied, their scaling
behavior is investigated and compared to
previous results, if available. In the last
section, a summary and an outlook are given.

\section{Algorithm}
\label{sec:algorithms}

The GS problem for general spin glasses is NP-hard
 \cite{SG-barahona82}. This means that only algorithms are known,
 where the running time in the worst case increases exponentially with
 the system size \cite{garey1979}. 
Hence, only small systems can be studied in the general case.
For the special case 
of two-dimensional spin glasses  without an external field and with
periodic boundary conditions in at most  one direction, i.e. for
planar graphs, efficient
polynomial algorithms \cite{SG-barahona82b} 
for the calculation of exact GSs 
are available. 
If one is interested in obtaining the partition sum, without obtaining
spin configurations, one can
also treat systems with full periodic boundary conditions in polynomial
time, by using transfer-matrix approaches 
\cite{saul1993,saul1994,regge1996,cho1997}, but the running
time is again strongly increasing, limiting the investigations to small
systems. 
The most recent studies are based on matching algorithms 
\cite{SG-bendisch1991,SG-bendisch1992,SG-bendisch1994,SG-bendisch1995,
SG-kisker1996,SG-kawashima1997,SG-bendisch1997,SG-bendisch1998,
SG-achilles2000}, \index{matching algorithm}
\index{algorithm!matching}
while other exact approaches can be found in
Refs. \cite{SG-ozeky1990,SG-kadowaki1997,middleton2001}. 
 Results for systems of size $1800 \times 1800$
have now been obtained \cite{SG-palmer1999}!

\begin{figure}[ht]
\begin{center}
\scalebox{0.44}{\includegraphics{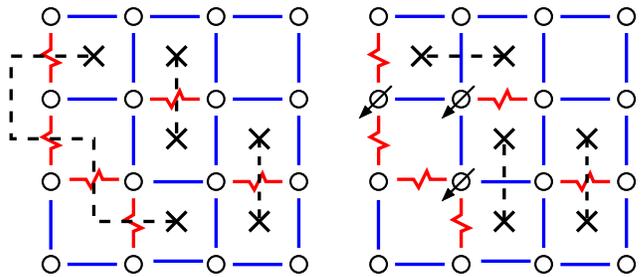}}
\caption{2d spin glass with all spins up (left, up spins not
  shown). Straight lines are ferromagnetic, jagged lines
  antiferromagnetic bonds. The dotted lines
  connect frustrated plaquettes (crosses). 
The bonds crossed by the dotted lines
are unsatisfied. In the right part the GS with three spins
pointing down (all other up), 
corresponding to a minimum number of unsatisfied bonds, is shown}
\label{fig:matching}
\end{center}
\end{figure}

Here, just the basic idea of the matching algorithm will be explained.
 For the details, see
Refs. \onlinecite{bieche1980,SG-barahona82b,derigs1991}. 
The method works for spin glasses which
are planer graphs. In the left part of
Fig.\ \ref{fig:matching} a small 2d system with open boundary
conditions is shown. All spins are assumed to be ``up'', hence all
antiferromagnetic bonds are not satisfied. If one draws a dotted line
perpendicular to all unsatisfied bonds, one ends up with the
situation shown in the figure: all dotted lines start or end at
frustrated plaquettes and each frustrated plaquette is connected to
exactly one other frustrated plaquette. Each pair of
plaquettes is then said to be {\em matched}. Now, one can consider the
frustrated plaquettes as the vertices and all possible pairs of
connections as the edges of a (dual) graph. 
The dotted
lines are selected from the edges connecting the vertices and called
a {\em perfect} matching, since {\em all} plaquettes are matched. 
One can assign the edges in the
dual graph weights, which are equal to the sum of the absolute values of the
bonds crossed by the dotted lines. The weight $\Lambda$
of the matching is defined as the
sum of the weights of the edges contained in the matching. As we have seen,
$\Lambda$ measures the broken bonds, hence, the energy of the configuration
is given by $E=-\sum_{\langle i,j\rangle} |J_{ij}|+2\Lambda$. Note
that this holds for {\em any} configuration of the spins, since a
corresponding matching always exists. Obtaining a GS means 
minimizing the total weight of the broken bonds (see right panel of
Fig.\ \ref{fig:matching}), so one is looking
for a {\em minimum-weight perfect matching}. This problem is solvable
in polynomial time.
The algorithms for minimum-weight perfect matchings 
\cite{MATCH-cook,MATCH-korte2000} are
among the most complicated algorithms for polynomial problems.
Fortunately the LEDA library offers a very efficient implementation
\cite{PRA-leda1999},  except that it consumes a lot of memory, which
limits the size of the systems to about $N=500^2$ on a typical 500 MB
workstation. 

For system with periodic boundary conditions in both directions, the
matching approach is not feasible. Hence, we restrict the study to
systems with free or with half periodic-half free boundary conditions.

To study the behavior of two-dimensional spin glasses, here not only
GSs, but also low-lying excited states are generated. For this
purpose GS algorithms can be used as well. The general
procedure consists of these  three steps:

\begin{enumerate}
\item Calculate the GS $\{S_i^{(0)}\}$ of a given
  realization and the GS energy  $E_0$.
\item Modify some of the couplings so that the GS is
  changed.
 \item Calculate the GS $\{S_i^{(m)}\}$ of the 
modified system, which is usually a
   low-lying excited state of the original realization. The 
   energy of $\{S_i^{(m)}\}$ calculated with the original (unmodified)
bonds is denoted by $E_0^m$.
\end{enumerate}

In this work four different ways of generating excitations are
considered. The technical details are in the corresponding 
sections.

\begin{itemize}
\item {\em Single-bond} droplets: the most straightforward implementation,
  where only one bond is changed with respect to the initial coupling.
\item {\em Cross droplets}: This mimics the generation of droplets induced
  by flipping one (central) spin. It involves changing $O(L)$ bonds,
  iterating over $2L-2$ changes, and picking the minimum-energy
  excitation among them.
\item {\em First excitations}: The lowest excitation above the GS is
  calculated. This involves the change of one bond (or  $O(L)$ bonds if
  the boundary spins are fixed), 
iterating over $O(L^2)$ changes, and picking the minimum-energy  
excitation among them.
\item {\em Fixed-volume} droplets: The matching algorithm does not
  allow for fixing the size. Hence, the size constraint is imposed
  here, when selecting among different {\em cross}
 droplets of each realization.
\end{itemize}

The main quantities analyzed are the energy $\Delta E = E_0^m-E_0$,
the volume $n$ and the surface $S$ of the droplets/excitations.

\section{Single-bond droplets}

\begin{figure}[htb]
\begin{center}
\myscaleboxB{\includegraphics{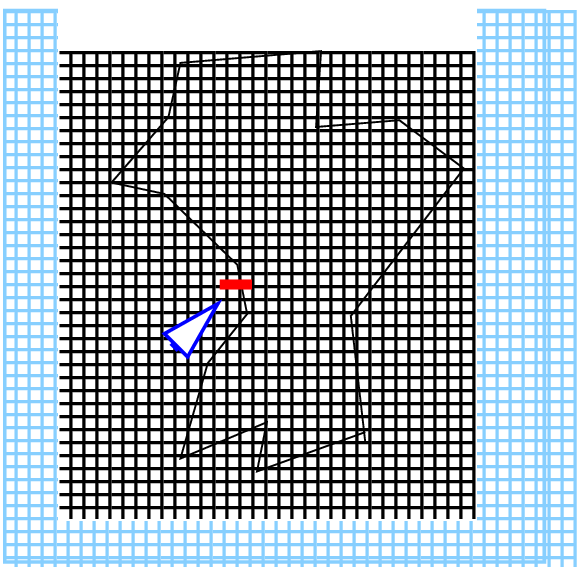}}
\end{center}
\caption{Method used to generate the {\em single-bond} droplets. 
After calculating the
  GS, one inverted hard bond (see triangle) generates
  an excitation (dark inner area).}
\label{figSingleBond}
\end{figure}

In Ref.\onlinecite{KA}, droplets were generated by flipping one
central spin with respect to the GS. Since matching
algorithms do not allow the inclusion of external fields, this
approach cannot be directly applied. The most natural and simple idea is
to flip instead a central {\em bond} with respect to the GS.
 This is performed by introducing a {\em hard} bond,
i.e. changing the value of one bond $(i_0,j_0)$ such that
$|J_{i_0j_0}|$ is so large $|J_{i_0j_0}|=J_{\rm big}$, 
that it will be always satisfied in a
GS, see Fig. \ref{figSingleBond}. 
In this case the hard bond $(i_0,j_0)$ will be {\em inverted},
i.e.  the spins $S_{i_0}$ and $S_{j_0}$ will be forced to take a relative
orientation opposite to the GS, hence 
$J_{i_0j_0}=-J_{\rm big}S_{i_0}S_{j_0}$. The GS of the
modified system now generates a minimum-energy droplet, called a {\em
  single-bond} droplet here, with the
constraint that the surface of the droplet runs through $(i_0,j_0)$.

\begin{figure}[htb]
\begin{center}
\myscalebox{\includegraphics{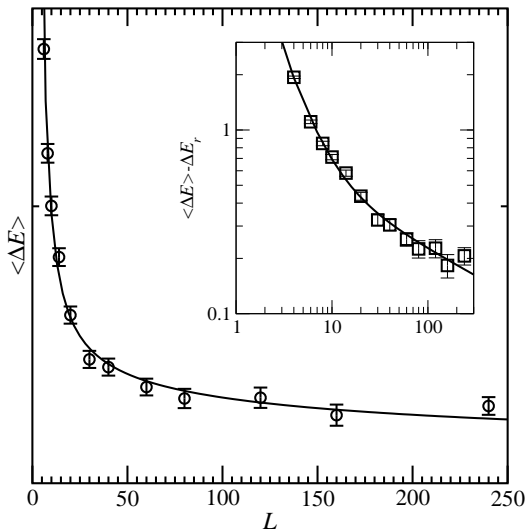}}
\end{center}
\caption{Average energy of {\em single-bond} droplets. The line shows
  a fit to the function $\Delta E(L)=\Delta
  E_r+AL^\theta+BL^{-\omega}$ with $\theta=-0.29$ resulting in
  $\Delta E_r =1.79(6)$ and $\omega=-1.8(2)$. 
The inset shows the same data with the
limiting energy $\Delta E_r$ subtracted in a double logarithmic plot.}
\label{figEDzero}
\end{figure}

We have studied {\em single-bond} droplets for system sizes $L=6$ to
$L=240$ while averaging over more than 3000 samples for each size.
In Fig.\ \ref{figEDzero} the average  droplet energy
$\langle \Delta E \rangle$
is shown as a function of the system size $L$. A double-logarithmic
plot (not shown) shows a curvature, which is not compatible with a
convergence of the droplet energy to zero, but compatible with
a convergence to a non-zero
energy. This seems plausible, because the choice that a certain bond,
i.e. the inverted hard bond in the modified system, has to
belong to the minimum domain wall, imposes a penalty on the energy of
the droplet. This energy should be of the order of the coupling
constant $J=1$. 
In the spirit of Ref. \onlinecite{droplets}, where a crossover of the
{\em cross} droplet energy 
according to  $AL^\theta+BL^{-\omega}$ has been observed,
we have performed a fit to a function  $\Delta E(L)=\Delta
  E_r+AL^\theta+BL^{-\omega}$ with fixing $\theta=-0.29$, resulting
 in  $\Delta E_r =1.79(6)$ and $\omega=-1.8(2)$, with a fair quality
 of the fit \cite{goodness}, $Q=0.14$. Interestingly, this is a similar
correction to scaling exponent $\omega$ as found when iterating the 
generation of
excitations over all single bonds and selecting the droplet from all
excitations containing the central spin \cite{droplets}.
One might be tempted to conclude that
 these types of droplet shows a similar behavior to the {\em
   cross} droplets studied in Ref. \onlinecite{droplets}, (see also the
 inset of Fig. \ref{figEDzero}).
But the fit is not very sensitive to the choice of $\theta$. If one
 allows also the variable $\theta$ to adjust in the fit, 
$\theta=-1.01(9)$, $\Delta
E_r=1.96(2)$ and $\omega=-6(4)$ results in $Q=0.83$.
Hence, this type of droplet seems to be very different
from the {\em cross}
droplets studied in Ref. \onlinecite{droplets}, (which are 
similar to the 
droplets generated by flipping a single spin within a 
Monte-Carlo minimization \cite{KA}).

\begin{figure}[htb]
\begin{center}
\myscalebox{\includegraphics{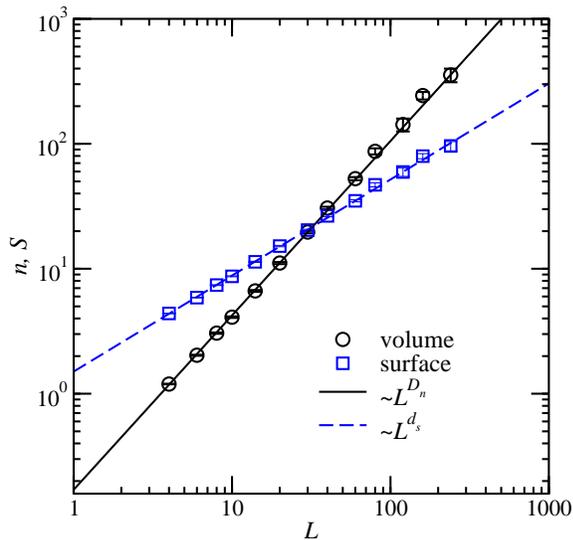}}
\end{center}
\caption{Average volume $n$ and average surface $S$ for the {\em
    single-bond} droplets as a function of the system size $L$. The
  straight lines represent fits to power-laws with exponents $D_n=1.39(1)$ 
resp.\ $d_S=0.769(5)$}
\label{figVSzero}
\end{figure}

This difference is visible also in the behavior of the {\em
  single-bond} droplet volume and surface, which are shown in Fig.
\ref{figVSzero}. We have fitted to power-law functions $L^{D_n}$ resp.\
$L^{d_S}$ resulting in exponents $D_n=1.39(1)$ 
and $d_S=0.769(5)$, hence $D_n/d_S=1.8$. This means this type of droplet
grows much slower than the size of the system, i.e. it is very different
from  growth as $L^2$ which was found for {\em cross} droplets
\cite{droplets}. 

Basically the above results mean that {\em
  single-bond} droplets are not the droplets which will dominate the 
low-temperature thermodynamics of the spin glass,
due to their high energy cost as  mentioned
above. Therefore, other procedures have to be applied to
obtain more physically relevant droplets when using a matching
 algorithm. However,
the single-bond droplets are not without their uses, 
as they are the first step
in obtaining the lowest energy excitations of the system, (see section V).

\section{Cross droplets}

The basic idea of the {\em cross} droplets is to mimic the approach,
which was used by Kawashima and Aoki to obtain droplets 
using a Monte Carlo simulation \cite{KA}. They first calculated the
GS. Then they recalculated the GS with the constraints
that the spins on the boundary keep their GS orientations
while a central spin is flipped. Using this approach, small system up
to $L=50$ could be studied, and a scaling $\Delta E \sim
L^{\theta^\prime}$ of the droplet energy with $\theta^\prime=-0.45(1)$
was found.

\begin{figure}[htb]
\begin{center}
\myscaleboxB{\includegraphics{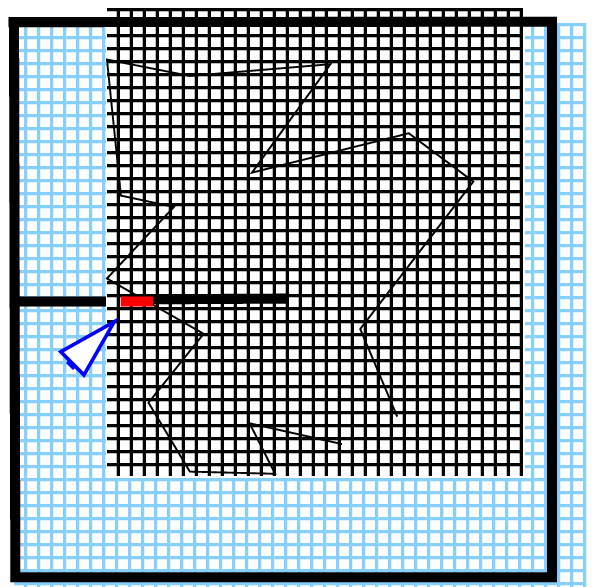}}
\end{center}
\caption{Method used to generate the {\em cross} droplets. 
After calculating the
  GS, several hard bonds (thick lines) are introduced, one
  hard bond is inverted (see triangle), leading to the appearance of an
  excitation (dark inner area).}
\label{figCrossDroplets}
\end{figure}

Our generation of the droplets works in the following way. After
obtaining the GS {\em several} hard bonds are introduced,
 see Fig. \ref{figCrossDroplets}.
If the subsystem of hard
bonds does not exhibit frustration, no hard bond will be broken, when a new
GS is calculated. First, all boundary spins are fixed relative to each
other by introducing hard bonds around the border,  i.e. 
the bonds between pairs of boundary spins are replaced. 
The sign of the hard bonds is chosen such that they are
 compatible with the GS orientations of the adjacent bonds. 
This keeps the spins on the boundary in their GS orientations.
Second, a line of hard
bonds is created which runs from  the middle of (say) the left border
to a pre-chosen center spin, 
again fixing the pair's spins in their relative GS orientations. 
Next, the sign of exactly one hard bond on this line is inverted. Finally,
a GS of the modified realization is calculated. With respect to
the original GS, the result is a minimum energy 
excitation fulfilling the constraints that it contains the center spin,
does not run beyond the boundary and that it has a surface which runs through
the hard bond which has been inverted. The energy of the excitation is 
defined as the
energy of the resulting configuration calculated using the original bond
configuration.
For each realization, this procedure is iterated over all the bonds which 
are located on the line from the boundary to the center, when in each case
exactly one hard bond is inverted. Furthermore,  this procedure is iterated
over all four lines of bonds running from the left, right, top and
bottom boundary to the center spin. Among all generated $2L-2$ excitations, the
one exhibiting the lowest energy is selected as the {\em cross} droplet.

In a previous work \cite{droplets}, we have found that the {\em cross}
droplets are almost equivalent to those of Kawashima and Aoki. Indeed for the
system sizes which can be studied, the difference is much smaller than
the statistical error bars. It turned out
that {\em cross} droplets exhibit for small sizes a
power-law behavior with an effective exponent $\theta^\prime=-0.47$.
But for large sizes a crossover to a power-law behavior is found,
governed by the
exponent $\theta=-0.29$, which is the same value
 as found in domain-wall studies 
\cite{bray1984,mcmillan1984b,rieger1996,palassini1999a,stiff2d,carter2002}.

Now we want to compare our results with recent results on fixed-volume
droplets obtained by Berthier and Young \cite{berthier2003}.
They have studied the energy
dependence of droplets for systems with $L=64$
when fixing the droplet size to an interval
$[n,n+\delta(n)]$. They considered two cases, $\delta(n)=0$ and
$\delta(n)=n-1$. For each realization, they optimized  the energy of a
droplet using a parallel-tempering Monte Carlo simulation. The
optimization was done while observing the constraints 
that the center spin is included, that the size is within
the given size range and that the droplet remains connected.
For both cases of $\delta(n)$, they observed for larger droplets a
scaling of the droplet energy which is compatible with the DS
scenario, i.e. a $\Delta E \sim n^{\theta^\star/d_f}$ behavior with
$\theta^\star=$ around -0.23 and $d_f=2$. For the case of fixed droplet
size ($\delta(n)=0$), this behavior was found also for the smallest
droplet sizes. For the fluctuating sizes ($\delta(n)=n-1$), a
crossover from the small-droplet behavior with a more negative
effective exponent to the large-droplet
behavior was found, similar to the crossover observed before for the
{\em cross} droplets \cite{droplets}.

Furthermore, they have studied the distributions $P(\Delta E)$
of droplet energies. In both cases they have found that the
distribution exhibits a scaling behavior according to
\begin{equation}
P_L(\Delta E) = \frac{1}{\langle \Delta E \rangle_L} P\left(\frac{\Delta
    E}{\langle \Delta E \rangle _L}\right)\,.
\label{eq:scaling:P}
\end{equation}
For the case $\delta(n)=0$ they found that $P_L(0)$ is zero (or very)
close, while for $\delta(n)=n$ a convergence to a non-zero value was
obtained. 

\begin{figure}[htb]
\begin{center}
\myscalebox{\includegraphics{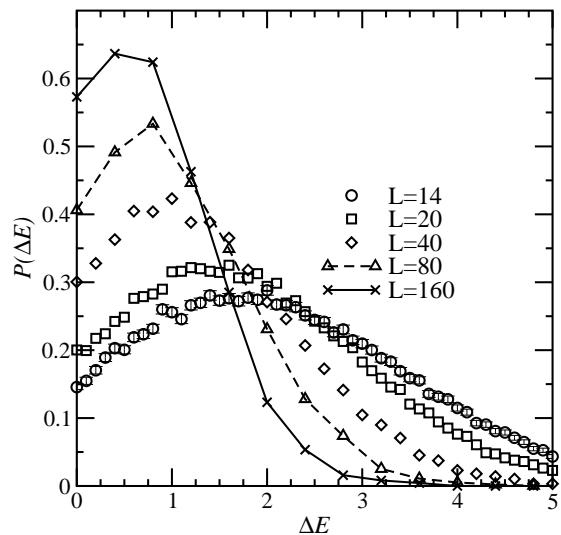}}
\end{center}
\caption{Distribution of droplet energies for {\em cross} droplets for
  system sizes $L=14,20,40,80,160$. Error bars are of the order of
  the symbol
  size (see $L=14$ datapoints). Lines are guides for the eyes only.}
\label{figCrossPE}
\end{figure}

\begin{figure}[htb]
\begin{center}
\myscalebox{\includegraphics{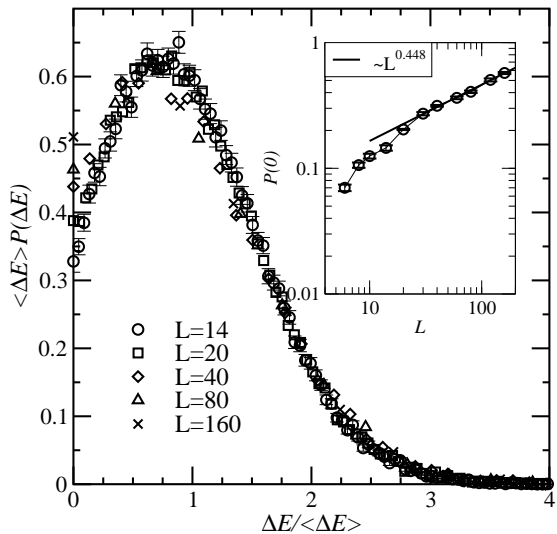}}
\end{center}
\caption{Rescaled distribution of droplet energies for {\em cross} droplets for
  system sizes $L=14,20,40,80,160$. The inset shows the probability
  $P(0)$ to get a zero-energy droplet as a function of system size
  $L$. For larger sizes, $P(0)$ scales as $L^{\theta_0}$ with
  $\theta_0=0.45(1)$.}
\label{figCrossPEScale}
\end{figure}

This difference between their two droplet types motivated us
to study $P(\Delta E)$ also for {\em cross} droplets.
In Fig.\ \ref{figCrossPE} the distributions of the droplet energies
for this droplet type is shown. 
We have studied system sizes up to $L=160$ and averaged over 5000
independent realisations in all cases.
One observes that $P(0)$ is
different from zero and growing with system size. The typical droplet
energy, i.e. the maximum of the distribution, is at a value different
from zero in the thermodynamic limit. 
This can be seen from  Fig.\ \ref{figCrossPEScale}, where
 the rescaled 
distributions $\langle \Delta E\rangle P(\Delta E)$ of the droplet energies
for the {\em cross} droplet type are shown. The scaling behavior of
$\langle \Delta E \rangle$ shows finite-size corrections, as discussed
in Ref.\ \onlinecite{droplets}. In that work,
 for large systems, $\langle \Delta E
\rangle\sim L^\theta$ with $\theta\sim - 0.28$ was found. Interestingly,
the scaling assumption (\ref{eq:scaling:P}) seems {\em not}
to work well near $\Delta E=0$, where a systematic increase of the
data points with system size is found, instead of a data collapse. To
investigate this effect, we have evaluated $P(0)$ as a function of
system size, see inset of Fig.\ \ref{figCrossPEScale}. Here again a
crossover can be observed. For larger sizes, we find $P(0)\sim
L^{\theta_0}$, with $\theta_0=0.45(1)$. Hence the scaling assumption 
$P_L(\Delta E)\sim \frac{1}{L^\theta}P(\frac{\Delta E}{L^\theta})$ may
be wrong.
But given the system sizes studied here, it cannot be
excluded that for even larger sizes, $P(0)$ shows another crossover to
the behavior found for the mean value $\langle \Delta E \rangle$
already at smaller systems.

To conclude, the scaling behavior of the distribution of droplet
energies for the cross droplets is similar to the $\delta(n)=(n-1)$
droplets of Berthier and Young, in the sense that $P(0)$ is finite and
growing. This similarity has been observed already, when studying
simply the mean value $\Delta E$. On the other hand, the scaling
assumption (\ref{eq:scaling:P}) seems not to work near $\Delta E=0$
in contrast to the results of Ref. \onlinecite{berthier2003}. Thus,
we have  another example of the scaling behavior of droplets seemingly
 depending
on the recipe used to generate them, at least at the system sizes which can 
be studied at present.


\section{First excitations}
\label{sec:lowest}

Next, we consider the first (i.e. the lowest) excitation of the system
 above the ground state.
They have been studied by Picco,
Ritort and Sales \cite{picco2001,picco2003} who generated the
lowest excitation for sizes up to $L=16$ using a transfer matrix
method. They have measured the exponent $\theta_1$ describing the 
finite-size scaling $\Delta E \sim L^{\theta_1}$ of the excitation
energy and the exponent $\lambda_l$ characterizing the size distribution
of the excitation volumes $n$ according to $P(n)\sim
n^{-(1+\lambda_l)}$. Furthermore, they have derived a scaling relation
$\theta = \theta_1 + d\lambda_l$ ($d=2$ here), connecting the exponents to the
usual droplet exponent $\theta$. Picco et al. argue $\theta_1=-2$
(while their numerical results are close to $\theta_1=-1.7(1)$, but
they explain this discrepancy as due to the  small sizes they could  study)
and they have found $\lambda_l=0.77(1)$, leading to $\theta=-0.46(2)$.
 
\begin{figure}[htb]
\begin{center}
\myscalebox{\includegraphics{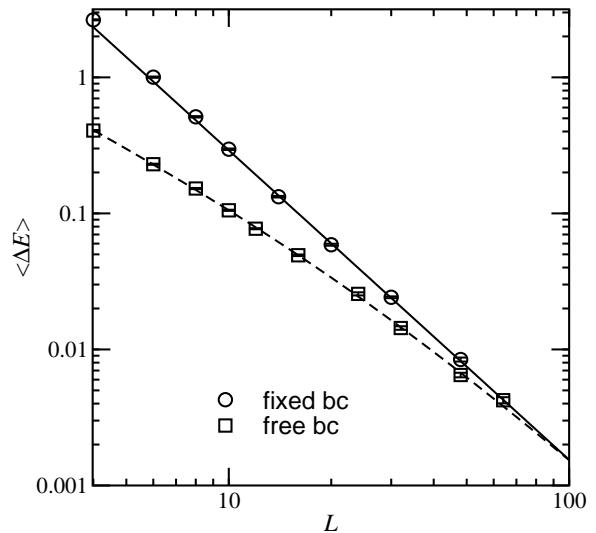}}
\end{center}
\caption{Average excitation energy $\Delta E$ as a function of system
  size for lowest excitations with free and with fixed boundary
  conditions. The lines denote fits to functions of the form $\Delta E_{\rm
    fixed}(L) = a L^{\theta_1}$ with $a=55(4)$, 
$\theta_1=-2.27(2)$ (fitted for $L\ge 10$) 
and $\Delta E_{\rm free}(L)=c L^{b'}/(1+eL^{b'-\theta_1})$ with $c=2.8(2)$,
$b'=-1.27(5)$, $e=0.041(3)$.}
\label{figFullPiccoE}
\end{figure}

As we have already seen for the single-spin induced (i.e. here {\em cross})
droplets, large sizes may be needed to see the correct scaling
behavior, hence $L=16$ may be much too small.
To test their results for larger sizes up to $L=64$ we have also
calculated first excitations by using the matching algorithm.
They are obtained by
iterating one hard bond over {\em all} bonds and then selecting for each
realization the minimum excitation among all $O(L^2)$ droplets.
We have considered free and fixed boundary conditions. In the latter
case, again a closed loop of hard bonds winds around the system.
We averaged over a number of realizations ranging from a few hundred
for the largest sizes to $10^4$ for $L\le 12$.
In Fig.\ \ref{figFullPiccoE} the average energy of the lowest
excitations for  both  types of boundary conditions 
are shown. For free boundary
conditions excitations leaving the boundary spins unchanged can be
realized. Thus, for each system, the energy of the lowest excitation with
fixed boundary conditions is an upper bound for the energy of the
excitation with free boundary conditions. Furthermore, in the
thermodynamic limit, the boundary conditions are expected to be
irrelevant for the energies of a droplet, hence both cases should
agree for $L\to\infty$. Indeed in Fig. \ref{figFullPiccoE}, the
average energy for the fixed boundary conditions is always above the
average energy for droplets with free boundaries. Furthermore, both
curves approach each other with growing system size. 
For small system sizes, the first excitations with free boundary conditions
can take advantage of the system boundary, i.e. many of the droplets
will run up to the boundary, where creating a domain wall does not
cost energy. Hence large corrections to scaling can be expected, as is
visible in the figure. In Ref. \onlinecite{picco2001} only small
systems $L\le 12$ were considered. Hence, the corrections to scaling
were hardly visible and a $L^{-1.7}$ behavior was observed.

Droplets for the case with
fixed boundaries show less corrections to scaling, because the
system cannot create domain walls of low energy cost.
 For $L\ge 10$ a
fit to an algebraic decrease of the excitation energy of the form
$L^{b}$ yields $b=-2.27(2)$.
Since the lowest excitations for both boundary conditions
 have to agree for $L\to\infty$ and because the
droplets with fixed boundaries exhibit almost no corrections to
scaling, it can be expected that also the lowest excitations with free
boundaries show the $L^{b}$ scaling for large sizes. This can be
represented by a function with a $L^{b'}/(1+eL^{b'-b})$ scaling
behavior, see Fig. \ref{figFullPiccoE}. But the system sizes
accessible to us, 
although much larger than those of Ref. \onlinecite{picco2001}, are 
still too small
to give a definite answer for this type of droplet.

\begin{figure}[htb]
\begin{center}
\myscalebox{\includegraphics{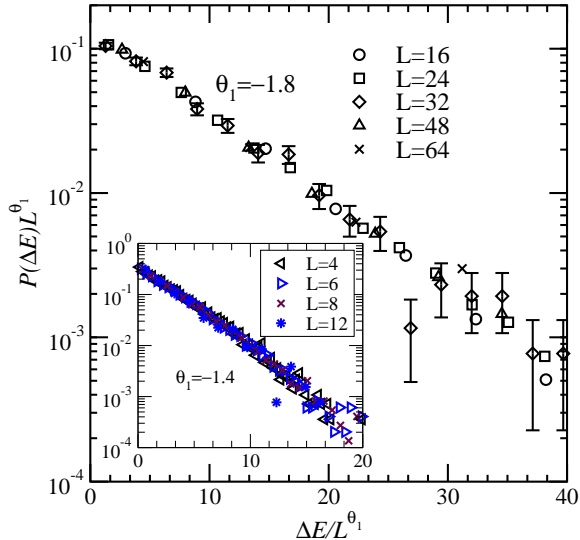}}
\end{center}
\caption{Rescaled distributions $P(\Delta E)L^{\theta_1}$ as a function of
the rescaled excitation energies $\Delta E/L^{\theta_1}$ for the
lowest excitations for
free boundaries. For large system sizes $\theta_1=-1.8$ was used, while
for small systems (see inset) $\theta_1=-1.4$ was applied. As an
example, the error bars for $L=32$ are shown, which are large in the
region of large, hence rare, energy values.}
\label{figPiccoDistr}
\end{figure}

The fact that  the droplets with free boundaries have larger 
finite-size corrections 
can be seen as well when studying the distribution of the droplet
energies. In Fig. \ref{figPiccoDistr} the rescaled distribution
$P(\Delta E)L^{\theta_1}$ of the droplets is displayed. For small systems
a good data collapse is obtained for $\theta_1=-1.4$, corresponding to
the slope of the mean droplet energy in Fig.\ \ref{figFullPiccoE}. For
larger sizes, the best data collapse is obtained for $\theta_1=-1.8$.

\begin{figure}[htb]
\begin{center}
\myscalebox{\includegraphics{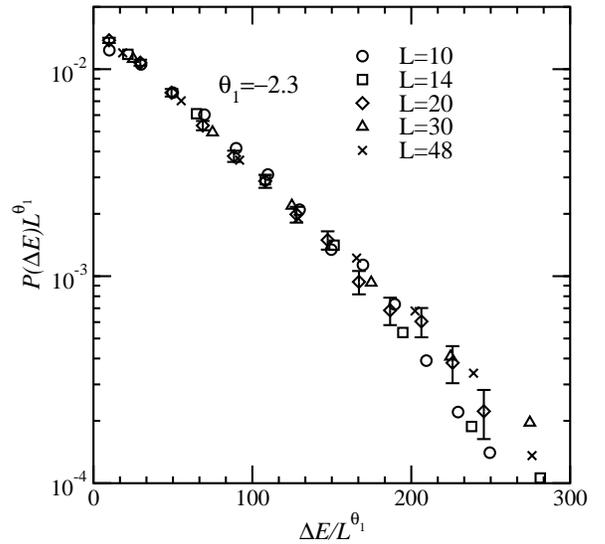}}
\end{center}
\caption{Rescaled distributions $P(\Delta E)L^{\theta_1}$ as a function of
the rescaled excitation energies $\Delta E/L^{\theta_1}$ for the lowest
excitations  for
fixed boundaries. $\theta_1=-2.3$ is used. As an
example, the error bars for $L=20$ are shown.}
\label{figFullDistr}
\end{figure}

For the case of {\em fixed} boundaries,
corresponding to the Fig. \ref{figFullPiccoE}, the best data collapse 
$P(\Delta E)L^{\theta_1}$ versus $\Delta E/L^{\theta_1}$ 
is obtained for $\theta_1=-2.3$, see Fig.\ \ref{figFullDistr}.

\begin{figure}[htb]
\begin{center}
\myscalebox{\includegraphics{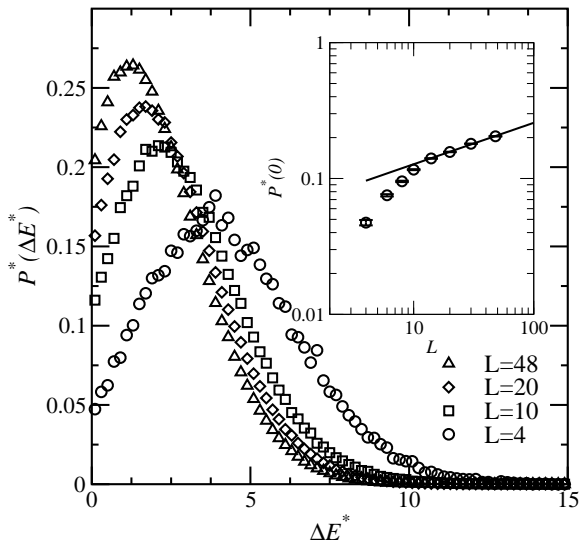}}
\end{center}
\caption{Distribution $P^\star(\Delta E^\star)$ of single-bond excitations
  energies with fixed boundary condition when
  iterated over all bonds of each realization, for system sizes
  $L=4,10,20,48$.  The inset shows the
  finite-size behavior of $P^\star(0)$. The line in the inset represents a
  $L^{0.3}$ behavior.
}
\label{figFullAllDistr}
\end{figure}

The different values for $\theta_1$ for the $L^{\theta_1}$ 
scaling can be understood as follows. The lowest
excitation is generated by iterating one hard bond over all bonds of a
system. Then the minimum-energy droplet is taken among all this
$O(L^2)$ excitations. To the lowest order, this procedure generates 
droplets, where the energy scales like 
\begin{equation}
\Delta E\sim 1/(L^2P^\star(0))\,,
\label{eq:scaling:deltaE}
\end{equation}
here $P^\star(0)$ is the probability
to obtain a (single-bond)
excitation having zero energy. Hence, to understand the scaling of $\Delta
E$, one should study the scaling of $P^\star(0)$.
In Fig. \ref{figFullAllDistr} the
distribution of energies of all single-hard-bond excitations (fixed boundary
conditions) is displayed along with
the finite-size behavior of $P^\star(0)$. For large enough sizes $P^\star(0)$
scales about as $L^{0.3}$ leading to $\Delta E \sim L^{-2.3}$ as found in
the data. Hence, the scaling  $P^\star(0)$ for the single-bond excitations 
is compatible with a $P^\star(0)\sim 1/L^\theta$ behavior with
$\theta\approx -0.3$, yet again the value found from domain wall studies.

Note that for small sizes, $P^\star(0)$ grows more rapidly than at larger
sizes, while from the scaling behavior of $\langle \Delta E \rangle$,
the opposite would be expected. This deviation is presumably
 due to the fact that for small sizes the behavior of the
energy distribution a little bit away from $\Delta E=0$ contributes as
well. In this region it was assumed  when deriving Eq.\
(\ref{eq:scaling:deltaE}), that the distribution is  constant, so
($P^\star(\Delta E^\star) \sim P^\star(0)$) for
small energies, which is, as visible in Fig.\ \ref{figFullAllDistr},
 {\em not} the case. It is only 
for larger sizes that the scaling behavior
{\em at} $\Delta E=0$ is relevant, leading to the expected result.

\begin{figure}[htb]
\begin{center}
\myscalebox{\includegraphics{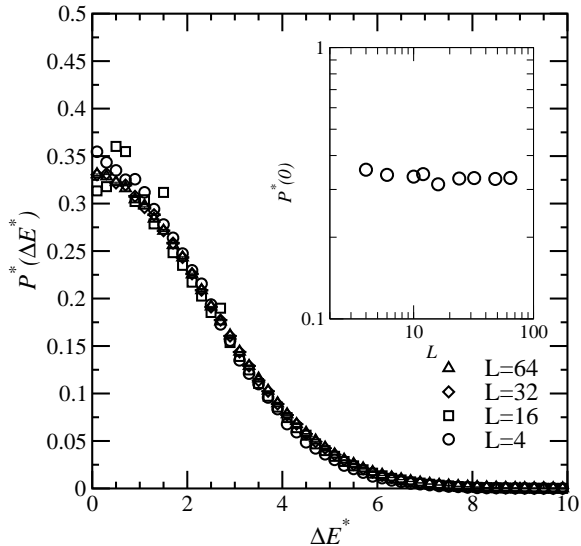}}
\end{center}
\caption{Distribution $P^\star(\Delta E^\star)$ of single-hard-bond excitations
  energies with free boundary conditions when
  iterated over all bonds of each realization, for system sizes
  $L=4,16,32,64$.  The inset shows the
  finite-size behavior of $P^\star(0)$. 
}
\label{figPiccoAllDistr}
\end{figure}

In Fig. \ref{figPiccoAllDistr} the corresponding distribution of
energies for the excitations with free boundary conditions
is shown. As one can see, excitations having zero energy are much more
likely than in the fixed boundary case, because they can take
advantage of letting the domain wall end at the boundary.
  For small sizes $P^\star(0)$ is
slightly decreasing, corresponding to the moderate decrease of
$\langle \Delta E \rangle$ seen in Fig. \ref{figFullPiccoE}, while for
moderate sizes, $P^\star(0)$ is constant as a function of system-size,
indeed compatible with $\theta_1=-2$, as argued by Picco et al.
Note that here the scaling of $\langle \Delta E \rangle$ is
also for small sizes compatible with the scaling of $P^\star(0)$, because
here $P^\star(\Delta E^\star)$ is much more constant near $\Delta E^\star=0$
and $P^\star(0)$ is much larger.

 For larger system sizes, when the system boundary is far away, the behavior
of $P^\star(0)$ for the two cases of boundary conditions should agree again. 
Unfortunatly, although the matching algorithm is a very powerful tool,
these sizes are out of reach with current technology. Hence it is hard
to say, whether $P^\star(0)\sim L^{0.3}$ or $P^\star(0)$ constant is the true
limiting behavior

\begin{figure}[htb]
\begin{center}
\myscalebox{\includegraphics{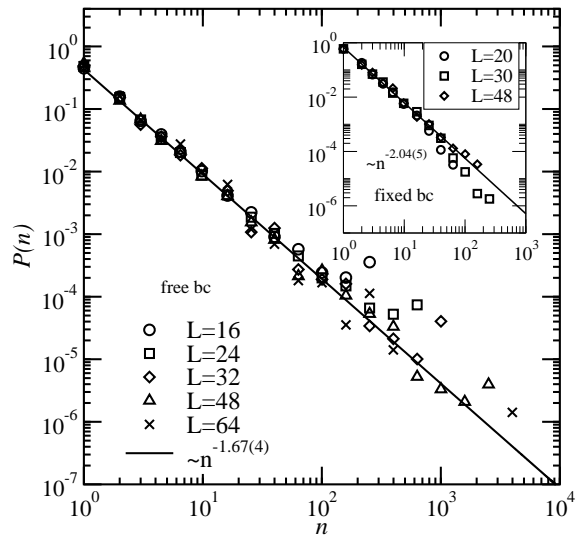}}
\end{center}
\caption{Distributions $P(n)$ of excitation volumes for lowest excitations
with free boundary conditions for different system sizes
$L=16,24,32,48,64$. Apart from the largest volumes, where the
excitations start to touch the boundary, the data is very well
described by a power-law behavior $n^{-1-\lambda_l}$ with
$\lambda_l=0.67(4)$ from a fit to the $L=48$ data in the range $n\in
[1,1000]$. The inset shows the same distribution for the case of fixed
boundary conditions. The line represents in this case a power law
with $\lambda_l=1.04(5))$.}
\label{figPiccoPn}
\end{figure}

Finally, to complete our comparison with the results of
Ref. \onlinecite{picco2003}, we show in
Fig. \ref{figPiccoPn} the distributions of excitation volumes for
the lowest excitations for both types of boundary conditions. It
exhibits a $P(n)\sim n^{-(1+\lambda_l)}$ behavior
\cite{picco2003}, except for the largest sizes, where the excitations
start to interact with the system boundary. We have fitted power-laws
to our data, ignoring the data points for the largest sizes and
obtained $\lambda_l=0.67(4)$ for free bc and $\lambda_l=1.04(5)$ for fixed bc. 
The former value is compatible with the previous
result $\lambda_l\approx 0.7$ obtained from direct finite-size scaling 
\cite{picco2003}, while the final value quoted in this reference is
$\lambda^{\rm eff}=0.77$ obtained from an aspect-ratio analysis,
i.e. using non-square systems. 

Anyway, it seems strange that
in both cases no large size-dependence of $\lambda_l$ is visible. Since
one would again expect that in the thermodynamic limit results obtained from
both types of
boundary conditions should agree, one would expect that the behavior
of one of the distributions should converge to the behavior of the
other. 

To summarize, for the behavior of the mean value
$\langle \Delta E\rangle$, the free-boundary case exhibits a
strong size dependence, while the fixed-boundary case exhibits almost
none. Hence in this case one would expect that for large sizes
the free-boundary case seems converged to the fixed-boundary results.
When studying $P^\star(0)$, {\em both} cases show a crossover, hence it
seems possible that one or both exhibit another crossover, leading to
the same thermodynamic limit. Finally, for the distribution of
sizes, it is hard to imagine that either shows a strong
crossover. There is a possibility that the
boundary conditions may indeed matter, and that a convergence to the same
behavior may {\em not} take place. 

\section{Fixed-volume droplets}

\begin{figure}[htb]
\begin{center}
\myscalebox{\includegraphics{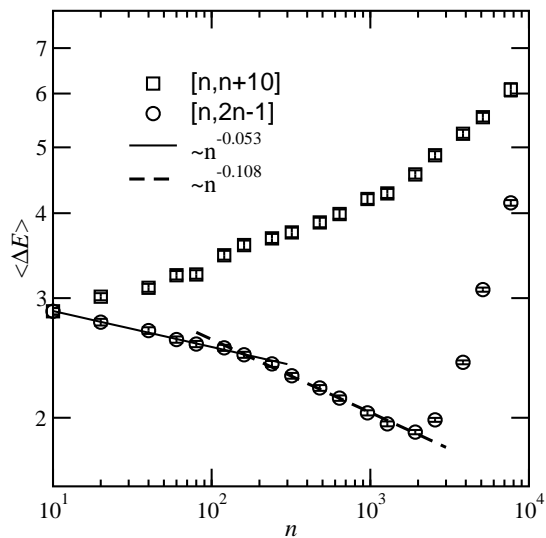}}
\end{center}
\caption{Average droplet energy for {\em cross} droplets ($L=160$), 
when the selection of the minimum droplet is restricted to a
excitation size
interval $[n,n+\delta(n)]$. Two cases $\delta(n)=9$ and
$\delta(n)=n-1$ are shown, in the spirit of the droplet generation
mechanism in Ref. \onlinecite{berthier2003}
}
\label{edCrossL160n}
\end{figure}

To compare with the results of Berthier and Young \cite{berthier2003}
we have implemented
a study similar to theirs for the size dependence for the {\em cross}
excitations. For the largest system size $L=160$, 
for each realization and a given size window
$[n,n+\delta(n)]$, the minimum-energy droplet was selected only among
the {\em cross} excitations which lie inside the size window. Here we have
studied also the case $\delta(n)=n-1$. Furthermore, instead of taking
$\delta(n)=0$, we have taken $\delta(n)=9$ to improve the statistics.

In Fig. \ref{edCrossL160n} the resulting average droplet energies are
shown. However, the behavior we
find here is totally different from that reported in
Ref. \onlinecite{berthier2003}. For the window of fixed length, the
average droplet energy even goes up with $n$, while for the case
$\delta(n)=n-1$, the droplet energy decreases, but for small sizes 
with a slower  $n^{-0.53}$ power-law decrease than the $n^{-0.22}$ 
decrease of
Ref. \onlinecite{berthier2003}. 
For larger droplet sizes, a crossover can again be observed, and the
data is compatible with a $n^{-0.11}$ decrease, which compares well
with the $n^{-0.115}$ behavior found in Ref.\onlinecite{berthier2003}.
For very large sizes $n$, the excitations touch the boundary of the
systems, where the spins are fixed. This leads to a strong increase of
the droplet energy.

To understand our result for small and medium droplets, one should first note
that there is an important difference between the droplets beind studied here
and in Ref. \onlinecite{berthier2003}. There, first the size was
fixed, then the optimization was performed 
while obeying the size
constraint. Here, first for each given inverted hard bond, a {\em
  first} optimization is performed, with a restricted search space. 
Then the sizes of the resulting
excitations are measured and finally the minimum-energy droplet is
selected among the excitations having the wanted sizes, i.e. a {\em second}
optimization is performed. Note that the matching algorithm
does {\em not} allow for fixing the size of a droplet in advance, so one
cannot circumvent the two-stage optimization process.

We assume now that 
the behavior of $\Delta E$ can be understood in the following way. The
distribution of excitation energies $P_n(\Delta E^\star)$, obtained after
the first minimization procedure, is taken as input. For each size, the
final droplet is selected as minimum 
among a given number of excitations which we
denote as $M(n)$. Hence a behavior
$\Delta E\sim 1/(M(n)P_n(0))$ can be expected, as in the case of the
first excitation discussed in the last section. We shall now test this
assumption using the available data for the case $\delta(n)=n-1$. 

\begin{figure}[htb]
\begin{center}
\myscalebox{\includegraphics{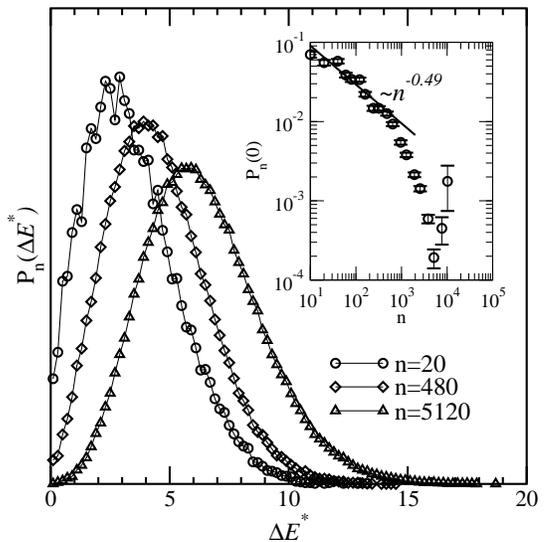}}
\end{center}
\caption{Distribution of energies for all {\em cross} excitations ($L=160$), 
restricted to a size interval $[n,n+\delta(n)]$, here
$\delta(n)=n-1$. The inset shows the behavior of $P(0)$ as a function
of size. The line in the inset denotes a power law $n^{-0.49}$ 
}
\label{edCrossL160Edistr}
\end{figure}

In Fig. \ref{edCrossL160Edistr} the distributions $P_n(\Delta E^\star)$
of excitation energies are shown for different sizes $n$. Also the
dependence of $P_n(0)$ on $n$ is shown in an inset. Due to numerical 
problems in determining $P_n(0)$, especially for small excitations, the
fluctuations are quite large. If one assumes a power law decrease, the
data is roughly compatible with a $n^{-0.49}$ behavior for small
$n$. But, in fact, no region with a clear power-law behavior can be easily
identified here.

\begin{figure}[htb]
\begin{center}
\myscalebox{\includegraphics{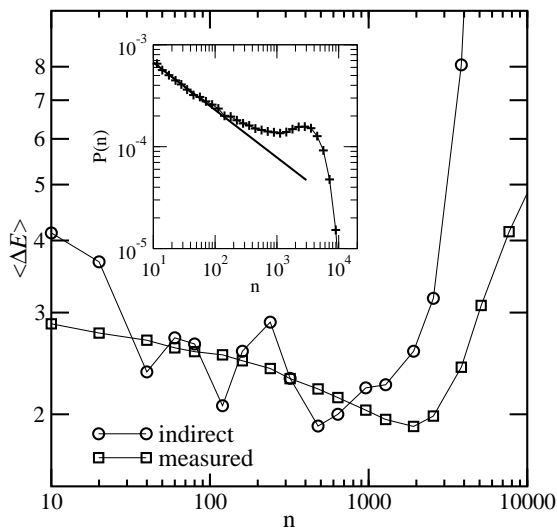}}
\end{center}
\caption{Average droplet energy for {\em cross} droplets ($L=160$), 
when the selection of the minimum droplet is restricted to a
excitation size interval $[n,2n-1]$. The numerical data from Fig. 
\ref{edCrossL160n} is repeated here (``measured'') 
and compared to the function
$c/(M(n)P_n(0))$, with the scaling factor $c=2700$ (``indirect''). 
The inset shows the distribution of sizes of the {\em cross}
excitations. $M(n)$ is obtained by integrating $P(n)$ over intervals
of size $\delta(n)$.
}
\label{edCrossL160Direct}
\end{figure}

$M(n)$ can be obtained from the distribution $P(n)$
of excitation sizes, see the inset of Fig. \ref{edCrossL160Direct},
via $M(n)\sim\int_n^{n+\delta(n)}P(n)dn$. In the main part of
Fig. \ref{edCrossL160Direct}, $c/(M(n)P_n(0))$, with $c=2700$
chosen by hand. is
compared to $\Delta E$.
The agreement is fair, and the large fluctuations originate in the
numerical problems when determining $P_n(0)$. 
One can circumvent this fluctuations for the small-size data, by
using scaling arguments. From a fit to the small size range $n\in[10,100]$ one
obtains $P(n)\sim n^{-0.46(2)}$. Since the selection is done in a
range $O(n)$, one has $M(n)\sim n^{-0.46+1}\sim n^{0.54}$. In combination with
the above result $P_n(0)\sim n^{-0.49}$ one obtains $\Delta E \sim
1/(M(n)P_n(0))=1/(n^{0.54}n^{-0.49})=n^{-0.05}$, in agreement with the
direct fit of the numerical data, see Fig. \ref{edCrossL160n}.

For
the case of fixed window size, each droplet is selected among
$M(n)\sim P(n)\sim n^{-0.46}$ droplets. Hence a increase of $\Delta E$
is not surprising here.
But the numerical problems due to a much smaller ($O(1)$ opposed to
$O(n)$) statistics when determining
$P_n(0)$ are too large in this case, such that a direct comparison
cannot be performed.

To summarize, the two recipes to generate fixed-size droplets are
different here and in the work of Berthier and Young. Hence, 
 in general different results come out. But for the case of
size fluctuations $\delta(n)=(n-1)$, the scaling behavior is the same.
 Since fluctuations $O(n)$ of the droplet
size seem to be the most natural case, directly in the spirit of the
droplet pictures by Fisher and Huse \cite{FH}, it seems not surprising that in
this case the results are similar. 
In our case it stems from a complex interplay of the selection
mechanism and the behavior of the distribution $P(n)$ of excitation
sizes as well as the distribution $P_n(\Delta E^\star)$ of excitation
energies for small energies. No direct relation to the scaling
exponent $\theta$ is obvious.
Furthermore, unfortunately, this still does
 not allow us to understand why for the $\delta(n)=O(1)$ case Berthier and
Young found the same scaling behavior, but without the
crossover usually seen at small to medium system sizes.

\section{Summary and outlook}

We have studied droplets and first excitations for two-dimensional
spin glasses with Gaussian interactions. Using highly sophisticated
matching algorithms, large sizes can be studied, but one is limited in
the flexibility of the recipes for droplet generation.

We studied four different types of droplets/excitations and observed
in general that the results depend strongly on the way droplets are
generated. The
{\em single-bond} droplets were very simple to generate. They
do not compare with typical droplets, which are assumed to be obtained
e.g. by the method of Kawashima and Aoki, because in the thermodynamic
limit their energy converges to a non-zero value.
The {\em cross} droplets, already introduced in Ref. \onlinecite{droplets},
are almost equivalent to the single-spin droplets, i.e. can be considered as
typical, and they show many properties expected from scaling
theory. Much larger sizes than with Monte Carlo methods can be
studied, hence the crossover to the $L^\theta$ ($\theta=-0.29$)
behavior observed. Also the distribution of the droplet energies is almost as
expected, except the fact the the scaling assumption does not work
very close to zero energy.
When studying {\em first excitations}, again much larger sizes than in
previous works are 
possible when using the matching approach. 
Different results are obtained for free and fixed boundary conditions.
Unfortunately, in this case the range of accessible sizes is even for
the powerful algorithm used still too small
to decide whether in the thermodynamic limit both cases agree and
what the final scaling behavior will be.
Finally, the study
of {\em fixed-volume} droplets shows the limitations of the matching
approach. It is not possible to use the matching algorithm to generate
droplets that are really equivalent to the droplets generated by
Berthier and Young. Nevertheless, it is possible to understand the
scaling behavior of the droplets generated here in terms of the
distributions of the excitations and their sizes. Furthermore, for the case of
fluctuating volumes, which is believed to represent the typical
behavior in the spirit of the DS theory, even the same scaling
behavior as obtained by Berthier and Young is found.

For the {\em single-bond} and the {\em cross} droplets, their behavior
seems to be clear.
For the {\em first excitations}, 
since the matching algorithm runs in polynomial time, and because the
generation of first excitations requires $O(L^2)$ calls, the
arrival of more powerful computers will resolve sooner or later
the question concerning the role of
boundary conditions.
It would be desirable to tackle also the last open point raised
by this paper: is it possible to extend or modify the matching
technique, such that real fixed-size droplets can be studied? This would
allow to understand their behavior better as well.
 
\begin{acknowledgments}
We would like to thank Marco Picco, Felix Ritort and Marta Sales for
providing some of their numerical data for comparison and
interesting discussions.
We are also indebted to Alan Bray for many discussions.
 AKH obtained financial support from the VolkswagenStiftung (Germany)
 within the program ``Nachwuchsgruppen an Universitäten''. Financial support
from the ESF and the Sphinx network is also acknowledged. 
 The simulations were performed at the ``Paderborn Center for
Parallel Computing'' and the ``Gesellschaft für Wissenschaftliche
Datenverarbeitung'' in Göttingen, both in Germany.  
\end{acknowledgments}

\end{document}